\documentclass[%
reprint,
nofootinbib,
 amsmath,amssymb,
aps,
pra,
]{revtex4-1}

\usepackage{graphicx}
\usepackage{dcolumn}
\usepackage{bm}
\usepackage{float}
\usepackage{pdfpages}

\usepackage{amsmath, amsthm, amssymb}

\newcommand{\be}{\begin{equation}}
\newcommand{\ee}{\end{equation}}
\newcommand{\bea}{\begin{eqnarray}}
\newcommand{\eea}{\end{eqnarray}}

\usepackage{rotating,wrapfig}
\usepackage{footnote}
\usepackage{upgreek}

\begin{document}


\title{Synthetic Landau Levels for Photons}
\author{Nathan Schine$^1$}
\author{Albert Ryou$^1$}
\author{Andrey Gromov$^2$}
\author{Ariel Sommer$^1$}
\author{Jonathan Simon$^1$}
\affiliation{$^1$ Department of Physics and James Franck Institute, University of Chicago, Chicago, IL}
\affiliation{$^2$Kadanoff Center for Theoretical Physics, University of Chicago, Chicago, IL}

\begin{abstract}
Synthetic photonic materials are an emerging platform for exploring the interface between microscopic quantum dynamics and macroscopic material properties\cite{Peyronel2012, Gopalakrishnan2009, Baumann2010, Cooper2013, Carusotto2013}. Photons experiencing a Lorentz force develop handedness, providing opportunities to study quantum Hall physics and topological quantum science\cite{Jia2015b, Otterbach2010, Wang2009}. Here we present an experimental realization of a magnetic field for continuum photons. We trap optical photons in a multimode ring resonator to make a two-dimensional gas of massive bosons, and then employ a non-planar geometry to induce an image rotation on each round-trip\cite{Yuan2007}. This results in photonic Coriolis/Lorentz and centrifugal forces and so realizes the Fock-Darwin Hamiltonian for photons in a magnetic field and harmonic trap\cite{Cooper2008}. Using spatial- and energy-resolved spectroscopy, we track the resulting photonic eigenstates as radial trapping is reduced, finally observing a photonic Landau level at degeneracy. To circumvent the challenge of trap instability at the centrifugal limit\cite{Cooper2008,Bloch2008}, we constrain the photons to move on a cone. Spectroscopic probes demonstrate flat space (zero curvature) away from the cone tip. At the cone tip, we observe that spatial curvature increases the local density of states, and we measure fractional state number excess consistent with the Wen-Zee theory, providing an experimental test of this theory of electrons in both a magnetic field and curved space\cite{Wen1992, Hoyos2012, Abanov2014, Can2014}. This work opens the door to exploration of the interplay of geometry and topology, and in conjunction with Rydberg electromagnetically induced transparency, enables studies of photonic fractional quantum Hall fluids\cite{Sommer2015, Umucalilar2014} and direct detection of anyons\cite{Paredes2001, Umucalilar2013}.
\end{abstract}

\maketitle

The Lorentz force on a charged particle moving in a magnetic field leads to the unique topological features of quantum Hall systems, including precisely quantized Hall conductance, topologically protected edge transport, and, in the presence of interactions, the predicted anyonic and non-abelian braiding statistics that form the basis of topological quantum computing\cite{Nayak2008}. To controllably explore the emergence of these phenomena, efforts have recently focused on realizing synthetic materials in artificial magnetic fields, and in particular, upon implementations for cold atoms and photons. Successful photonic implementations have employed lattices with engineered tunneling\cite{Jia2015b, Wang2008, Rechtsman2013, Hafezi2013, Rechtsman2013b}. However, it is desirable to realize artificial magnetic fields in the simpler case of a continuum (lattice-free) material\cite{Otterbach2010, Karzig2015, Longhi2015}, where strong interactions are more easily accessible and the theory maps more directly to fractional quantum Hall systems. In this work, we develop a new approach and demonstrate the first continuum synthetic magnetic field for light.

\begin{figure*}[t!]
\includegraphics[width=.6\linewidth]{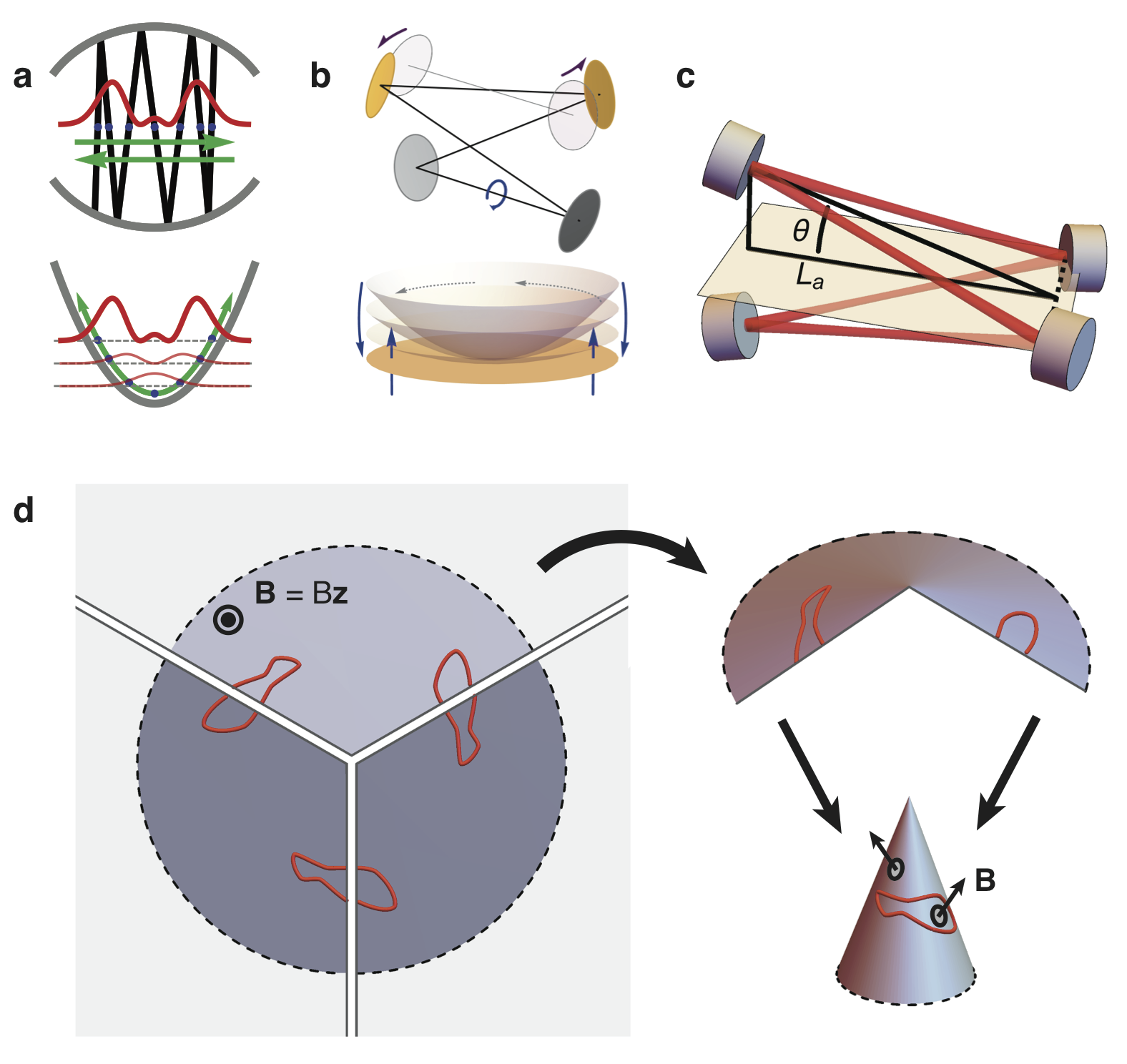}
\makeatletter
\renewcommand{\@makecaption}[2]{%
  \par\vskip\abovecaptionskip\begingroup\small\rmfamily
  \splittopskip=0pt
  \setbox\@tempboxa=\vbox{
    \@arrayparboxrestore \let \\\@normalcr
    \hsize=.5\hsize \advance\hsize-1em
    \let\\\heading@cr
    \@make@capt@title {#1}{#2}
  }%
  \vbadness=10000
  \setbox\z@=\vsplit\@tempboxa to .55\ht\@tempboxa
  \setbox\z@=\vtop{\hrule height 0pt \unvbox\z@}
  \setbox\tw@=\vtop{\hrule height 0pt \unvbox\@tempboxa}
  \noindent\box\z@\hfill\box\tw@\par
  \endgroup\vskip \belowcaptionskip
}
\makeatother
\caption{\textbf{Resonator structure and transverse manifold geometry. a,} Top, ray trajectories (black lines) in a curved mirror resonator oscillate transversely (green arrows). In a particular transverse plane, the stroboscopic time evolution of the ray-positions samples a harmonic oscillator trajectory (blue points). In paraxial optics, the solutions for the transverse modes are Hermite-Gauss profiles (red curve). The transverse degrees of freedom of a resonator are precisely those of a 2D quantum harmonic oscillator (below). \textbf{b}, Top, as a four mirror resonator is made non-planar (purple arrows), the light rays are induced to rotate (blue arrow) about the optic axis. In the transverse plane (represented below), this corresponds to flattening the 2D harmonic potential (centrifugal force) and the introduction of an effective magnetic field (Coriolis force). \textbf{c}, Our non-planar resonator consists of four mirrors (blue and purple) in a stretched tetrahedral configuration of on-axis length $L_a$ and opening half-angle $\theta$. The image rotates about the optic axis (red) on every round trip. \textbf{d}, We depict the transverse plane at the resonator waist pierced by a uniform perpendicular (along $\hat{\textbf{z}}$) magnetic field $\textbf{B}$ of magnitude $B$, and show a generic profile (red curve) with threefold symmetry. When the plane is cut arbitrarily into three equal sections, the entire profile is fully determined within any one-third section of the plane: when a trajectory leaves one side of a section, it reappears on the other side. Each section may be wrapped into a cone on which the original profile appears once (this would be true for any discrete rotational symmetry). The effective magnetic field is everywhere perpendicular to the cone's surface.}
\label{fig:Fig1}
\end{figure*}

\begin{figure*}[t!]
\includegraphics[width=.99\linewidth]{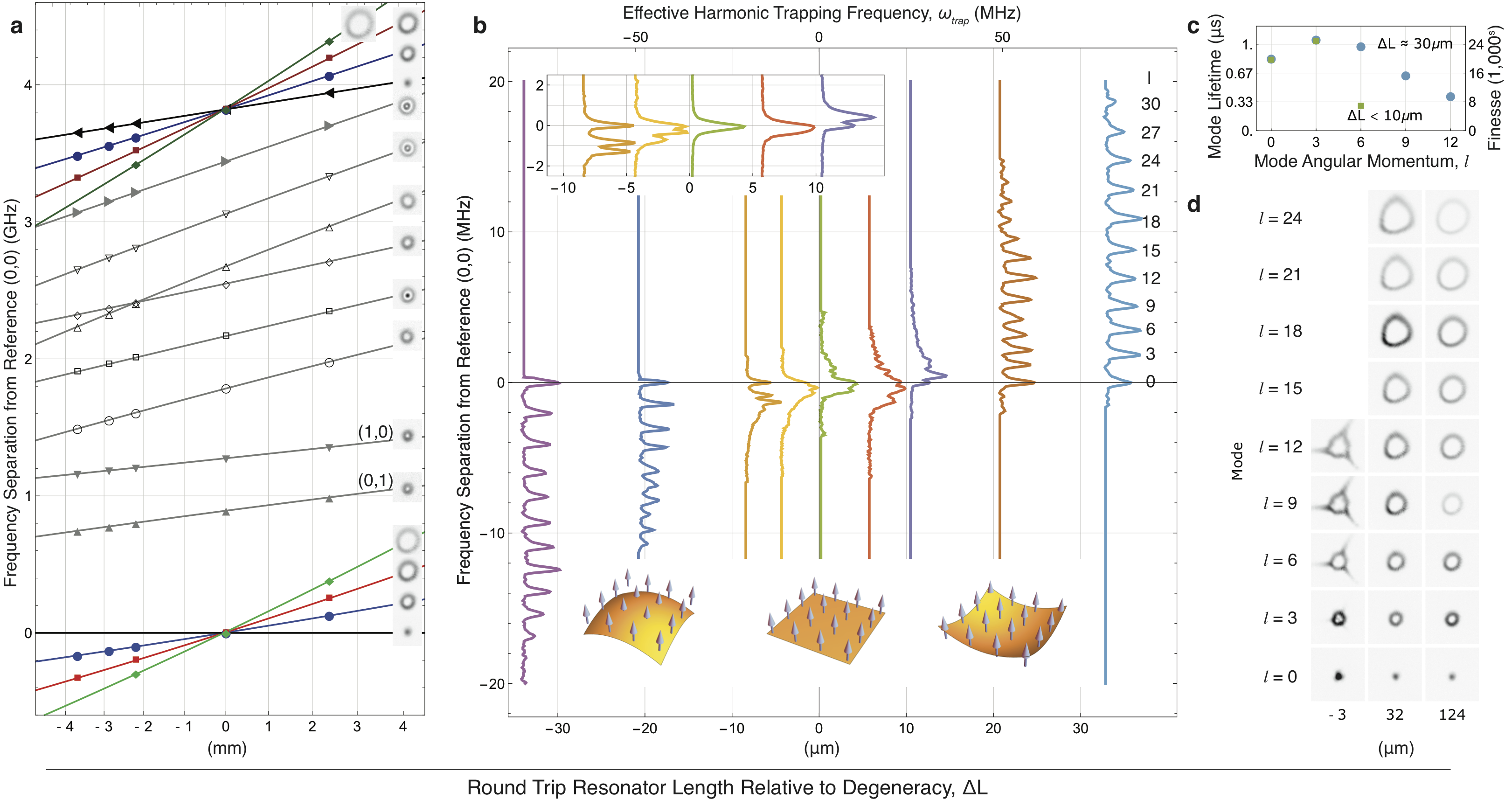}
\makeatletter
\renewcommand{\@makecaption}[2]{%
  \par\vskip\abovecaptionskip\begingroup\small\rmfamily
  \splittopskip=0pt
  \setbox\@tempboxa=\vbox{
    \@arrayparboxrestore \let \\\@normalcr
    \hsize=.5\hsize \advance\hsize-1em
    \let\\\heading@cr
    \@make@capt@title {#1}{#2}
  }%
  \vbadness=10000
  \setbox\z@=\vsplit\@tempboxa to .55\ht\@tempboxa
  \setbox\z@=\vtop{\hrule height 0pt \unvbox\z@}
  \setbox\tw@=\vtop{\hrule height 0pt \unvbox\@tempboxa}
  \noindent\box\z@\hfill\box\tw@\par
  \endgroup\vskip \belowcaptionskip
}
\makeatother
\caption{\textbf{Building a Landau level.} The modes of our resonator follow the Fock-Darwin Hamiltonian of a massive, harmonically trapped particle in magnetic field: the magnetic field creates a ladder of Landau levels uniformly spaced by the cyclotron frequency, $\omega_c$, while the harmonic trap of frequency $\omega_{trap}$ uniformly splits levels within each Landau level by $\omega_{trap}^2/\omega_c$ (see Supplementary Information). We probe this spectrum versus resonator length $L_{rt}$, and demonstrate that, for each $L_{rt}$, the spectrum is determined by two energies $\nu_{(1,0)}$ and $\nu_{(0,1)}$ according to $\nu_{(\alpha,\beta)}=\alpha \nu_{(1,0)}+\beta \nu_{(0,1)}  \mod \nu_{FSR}$, where $\omega_c = 2\pi \times \nu_{(1,1)}$ gives the cyclotron frequency and $\omega_{trap}^2/\omega_c = 2\pi \times \nu_{(3,0)}$ provides the harmonic trapping frequency. Furthermore, fine-tuning $L_{rt}$ drives $\omega_{trap}$ to zero, bringing specific sets of angular momentum eigenmodes into degeneracy, thereby forming Landau levels. \textbf{a,} The frequency separations between several modes and a reference $l = 0$ mode are plotted as the harmonic confinement is coarsely tuned relative to an approximately degenerate reference length $L_{rt} = 78.460$ mm (corresponding free spectral range $\nu_{FSR} = 3.8209$ GHz). Solid lines are obtained as integer linear combinations of fits to the modes labeled (1,0) and (0,1) and the free spectral range. For details on mode indexing, see Supplementary Information. \textbf{b,} Main panel, we plot the transmission spectrum of the first $\sim 10$ modes in the lowest Landau level against small deviations from nominal degeneracy. Top inset, low order modes become degenerate to within a resonator linewidth, $\kappa \approx 200$ kHz, while in the main panel, we observe weak level repulsion (approximately equal to the resonator linewidth) in the higher order modes consistent with mode mixing due to mirror imperfections of $\sim \lambda/5000$. $\omega_{trap}$ is presented on the upper horizontal axis. \textbf{c,} The lifetimes (and corresponding finesses) of representative modes decrease for higher mode numbers both away from degeneracy (blue circles) and near degeneracy (green squares). Here $\Delta L$ is the offset of the round trip resonator length from nominal degeneracy. \textbf{d,} With significant residual harmonic trapping ($\Delta L = 124$ $\mu$m), angular momentum modes are simple rings. As the trapping is reduced ($\Delta L = 32$ $\mu$m),  high angular momentum modes begin to mix due to local disorder. When the trapping is precisely cancelled ($\Delta L = -3$ $\mu$m), mirror imperfection consistent with a single nanoscopic scratch dramatically alters the modes’ shape away from the predicted near-Laguerre-Gauss profiles. Even the first resonator mode is noticeably triangular, indicating at least a mixing of Laguerre-Gauss $l = 0$ and $l = 3$ modes. Overcoming this disorder necessitates only $\sim$MHz photon-photon interactions to explore strongly correlated physics.}
\label{fig:Fig2}
\end{figure*}

To achieve photonic Landau levels we harness the powerful analogy between photons in a near-degenerate multimode cavity and massive, trapped 2d particles\cite{Klaers2010, Sommer2016}. Owing to mirror curvature, the transverse dynamics of a running wave resonator are equivalent to those of a 2D quantum harmonic oscillator (Fig. 1a). Non-planar reflections cause the transverse properties of the light field---for example, field profile (image) and polarization vectors---to rotate by an angle $\phi$ upon a round trip (Fig. 1b). Polarization rotation splits the energy of circularly polarized eigenmodes, while image rotation, in analogy to a rotating frame, introduces Coriolis and centrifugal forces. As the anti-confinement from the rotation compensates the confinement from the mirror curvature, we are left primarily with a Coriolis force, or equivalently, a Lorentz force. When dynamics are coarse-grained over many-round trips, we arrive at the Fock-Darwin Hamiltonian (see Supplementary Information) $H_{FD}=\frac{1}{2m} (\vec{\textbf{p}}-\frac{(qB)^{syn}}{2} \hat{\textbf{z}} \times \vec{\textbf{r}})^2+\frac{1}{2} m \omega_{trap}^2 r^2$, where $m$ is the dynamical particle mass, $\vec{\textbf{p}}$ is the particle's transverse momentum vector, $\vec{\textbf{r}}$ is the particles transverse position vector, $\hat{z}$ is the longitudinal unit vector, and $\omega_{trap}/2\pi$ is the (residual) harmonic trapping frequency. The synthetic magnetic field is\cite{Sommer2016} $(qB)^{syn}/\hbar =\frac{2\pi}{\lambda L_a} \phi \approx \frac{8\pi}{\lambda L_a} \theta^2$ for small angles $\theta$, where $L_a$ and $\theta$ are the on axis resonator length and opening half angle (Fig. 1c), and $\lambda$ is the wavelength of light. When the resonator length is tuned to eliminate residual harmonic trapping, only a Lorentz force remains, and the Hamiltonian describes massive particles in Landau levels, where the \emph{n}th Landau level has energy $\hbar \omega_c (n+\frac{1}{2})$, with $\omega_c$ being the cyclotron frequency, and consists of states with angular momentum $l = -n, -n + 1, \ldots$ in units of the angular momentum quantum $\hbar$. The synthetic magnetic field is then equivalently given by $(qB)^{syn}/\hbar=4/w_0^2$, that is, one flux quantum per area $\pi w_0^2/4$, where $w_0$ is the resonator $l = 0$ mode waist ($1/e^2$ intensity radius). The magnetic length $l_B$ may therefore be identified as $w_0/2$.

Although Landau levels exhibit `topological protection' against localized disorder, long-range potentials may guide the particles to infinity, inducing loss\cite{Bloch2008, Schweikhard2004}. In our system, the dominant source of long-range disorder is trap asymmetry (astigmatism) that arises from mirror imperfections and off-axis reflection and drives $\Delta l = \pm 2$ transitions (see Supplementary Information). We circumvent this by imposing an additional discrete threefold rotational symmetry on our Landau levels. To achieve this, we carefully balance transverse and longitudinal energy scales such that only every third angular momentum state is degenerate (see Supplementary Information).

The threefold symmetry of the Landau levels induces a conical geometry on the 2D space for transverse photon dynamics. To see this, consider a particle which leaves the edge of a particular $120^{\circ}$ wedge of the plane; the discrete rotational symmetry requires it to appear on the other side, which is equivalent to wrapping this wedge into a cone (Fig. 1d). Working away from the apex of the cone gives access to flat space Landau levels with every angular momentum state accessible, while working near the apex allows experimental investigation of particle dynamics near a singularity of spatial curvature. 

Our experimental resonator consists of four mirrors with nominal radii of curvature $R = (2.5, 5, 5, 2.5)$ cm arranged as shown in Fig. 1c, and has an $l = 0$ mode finesse of $2. \times 10^4$. The on-axis length $L_a = 1.816$ cm and the opening half-angle $\theta = 16^{\circ}$ were chosen to create a photonic Landau level while minimizing residual astigmatism. Varying the resonator length by $\sim 20$ $\mu$m adjusts the splitting between states by $\sim 1$ MHz (see Supplementary Information). Tuning this splitting to zero results in a free spectral range at degeneracy of $\nu_{FSR} = 3.8209(2)$ GHz. The resonator has an $l = 0$ waist size $w_0 = 43$ $\mu$m and a cyclotron frequency $\omega_c = 2\pi \times 2.1671(2)$ GHz, which together yield a photon dynamical mass of $m_{dyn} = \frac{4\hbar}{\omega_c w_0^2} = 1.84 \times 10^{-5}$ $m_e$ where $m_e$ is the electron mass. 
\begin{figure*}[t!]
\includegraphics[width=.5\linewidth]{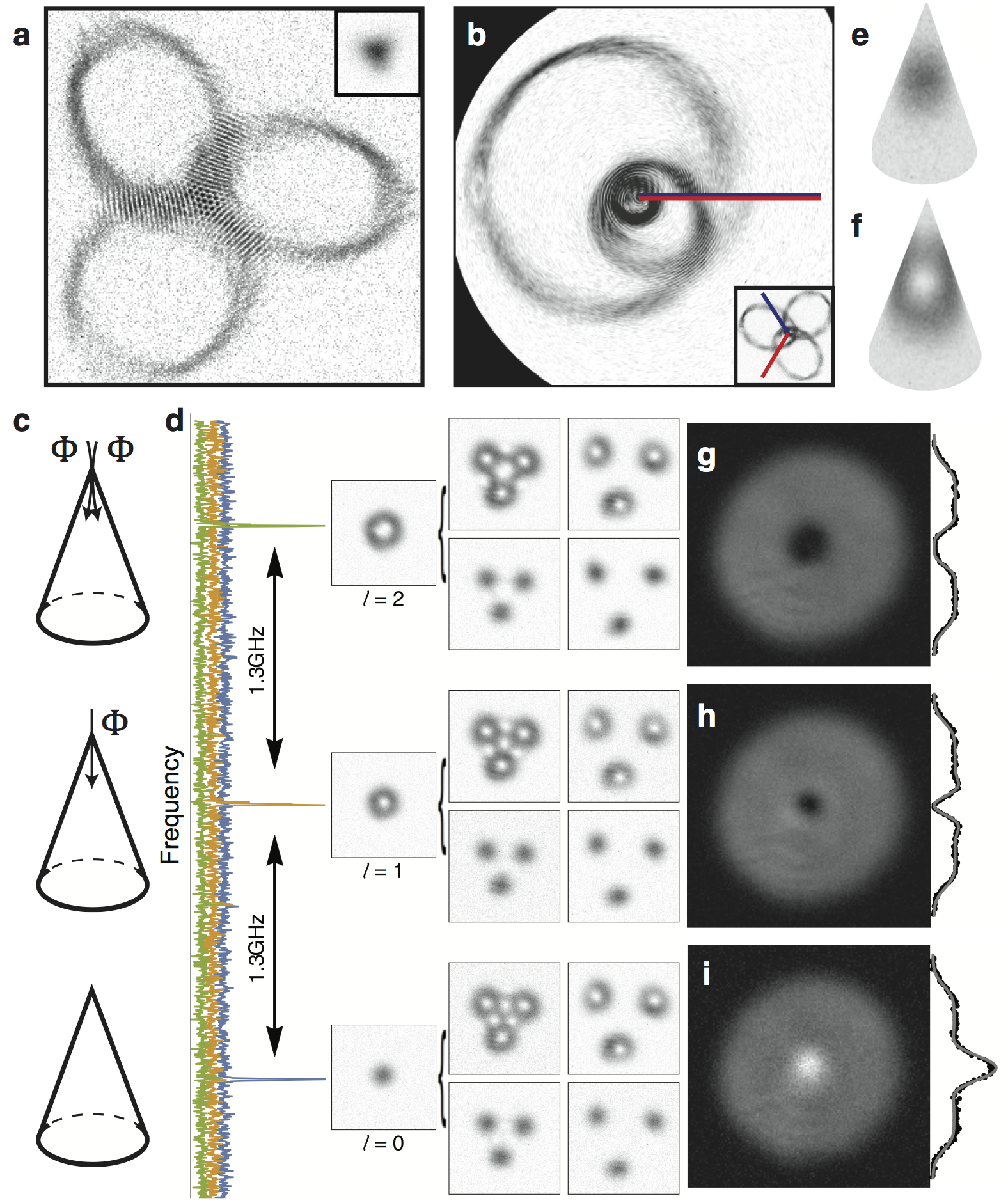}
\makeatletter
\renewcommand{\@makecaption}[2]{%
  \par\vskip\abovecaptionskip\begingroup\small\rmfamily
  \splittopskip=0pt
  \setbox\@tempboxa=\vbox{
    \@arrayparboxrestore \let \\\@normalcr
    \hsize=.5\hsize \advance\hsize-1em
    \let\\\heading@cr
    \@make@capt@title {#1}{#2}
  }%
  \vbadness=10000
  \setbox\z@=\vsplit\@tempboxa to .55\ht\@tempboxa
  \setbox\z@=\vtop{\hrule height 0pt \unvbox\z@}
  \setbox\tw@=\vtop{\hrule height 0pt \unvbox\@tempboxa}
  \noindent\box\z@\hfill\box\tw@\par
  \endgroup\vskip \belowcaptionskip
}
\makeatother
\caption{\textbf{Photonic lowest Landau levels on a cone. a,} At degeneracy, all resonator modes display three-fold symmetry. We present a very large displaced angular momentum mode with radial extent up to 8 times the mode waist, $w_0$, implying that $\sim$20 modes must be degenerate. The rapid phase winding for large $l$ modes causes the strong fringing pattern when the mode self-interferes. Inset, an $l = 0$ mode at the same scale. \textbf{b,} We project another large angular momentum mode onto a cone and view it from above the apex. We observe a general property that circular orbits must encircle the cone apex either zero or three times. Inset, the original image of the mode. The pair of rays overlaying the inset image corresponds to the cut in the main image. \textbf{c,} The twisted resonator corresponds to Landau levels on three cones with differing quantities of magnetic flux threaded through the tip. The cone built out of $l = 0, 3, 6, \ldots$ has no flux threading; the cone built out of $l = 1, 4, 7, \ldots$ is threaded by $\Phi_0/3$; and the cone built out of $l = 2, 5, 8, \ldots$ is threaded by $2\Phi_0/3$, where $\Phi_0$ is the magnetic flux quantum. \textbf{d,} With the resonator tuned to degeneracy, we identify the energies of the $l=c$ modes for $c = 0, 1,$ or $2$ by the transmission peaks (blue, orange, and green curves, respectively) that correspond to the correct observed transmitted mode's profile (single images, labelled). The degenerate sets starting with these modes each form a lowest Landau level on different cones. Except at the apex, each cone is flat, so away from the tip each lowest Landau level supports modes of---and therefore the dynamics of---a planar lowest Landau level with $l = 0, 1, 2, \ldots$ defined about a displaced point. On each cone, we show displaced $l = 0$ (bottom two) and $l = 1$ (top two) modes. For large displacements (right two), these modes are undistorted; however, for small displacements (left two), where there is significant mode amplitude at the tip, we observe distortions due to self-interference, similar to panel \textbf{a}. \textbf{e,f,} Displaced $l = 0$ and $l = 1$ modes from \textbf{d} are projected onto a cone to show how observed mode images may be interpreted on a conical surface.  \textbf{g-i,} We explore the effects of curvature and flux threading near the tip by measuring the local density of photonic states. For the $c = 0$ cone (\textbf{i}), we find an approximately threefold increase in local state density near the cone apex above a constant background plateau of density. This corresponds to an additional one-third of a state localized near the apex. For the cones with $c = 1$ and $2$ (\textbf{h} and \textbf{g}, respectively), we find a vanishing local density of states near the apex, reflecting the negative magnetic flux threading through the cone apex. Each unit of flux removes one-third of a state local to the apex so that the $c = 1$ cone has no additional states, and the $c = 2$ cone is missing one-third of one state. The data to the right display a slice through the middle of each image; the gray curves are fits to the expected analytic form (see Supplementary Information). }
\label{fig:Fig3}
\end{figure*}

In practice, we tune our resonator to degeneracy by varying its length, which primarily changes the harmonic trapping without changing the effective magnetic field, and we track the energy spectrum and spatial profiles of resonator modes by observing the transmission of circularly polarized light with a holographically programmed spatial profile (Fig. 2, see Supplementary Information). Figure 2a shows the evolution of a number of mode energies in numerous Landau levels as we adjust the resonator length over almost a centimetre. Using the observed mode-profiles (shown as insets), we identify the four lowest modes in the figure as those comprising the lowest conical Landau level, and centre the graph on their approximate degeneracy point. Figure 2b shows high-resolution spectroscopy of a larger number of modes in the lowest Landau level near the length where the harmonic confinement is cancelled. We precisely extract the change in resonator length from the spectroscopically measured free spectral range and compensate the residual harmonic trapping to zero. At this point, the residual non-degeneracy comes from local disorder, which causes an observed level repulsion for high angular momentum states (Fig. 2b, main panel) that is not observed at lower angular momentum (Fig. 2b, top inset) as well as a significant reduction in mode lifetime (Fig. 2c). Away from degeneracy the modes are nearly ideal rings with $2\pi \times l$ phase winding (experimentally determined by varying the phase profile of the injected light, see Supplementary Information); at degeneracy these modes mix due to local disorder potentials (Fig. 2d). This effect is apparent because of the long particle lifetime (high finesse of our resonator) and, in only causing mode distortion, is qualitatively different from global potentials such as astigmatism that cause mode deconfinement (see Supplementary Information). The local disorder merely creates chiral, localized states; it does not break topological protection so long as it only mixes modes within a single Landau level and, in an interacting system, is weaker than the interactions. This insensitivity to weak disorder is a notable advantage of our setup as compared to, e.g. injecting angular momentum modes into a two mirror resonator (see Supplementary Information). 

To demonstrate our system's stability out to large displacements from the cone tip, figures 3a-b show large angular momentum orbits. Figure 3a presents a large displaced state composed of modes with angular momentum up to $l \approx 60$, which exhibits threefold symmetry and interferes with itself, producing a strongly fringed pattern due to the rapid phase winding of each ring. Figure 3b unwraps another large angular momentum mode showing that if an orbit encircles the cone tip, then it must do so three times, as a consequence of the threefold symmetry. 

Remarkably, photons in our resonator may live on three distinct cones, distinguished by additional magnetic flux threaded through their tips. To understand this, note that the planar lowest Landau level may be spanned by angular momentum states $\psi_l (z=\frac{x+iy}{w_0}) \propto z^l  \exp⁡(-|z|^2)$ for $l = 0, 1, 2, \ldots$, with the transverse position vector $\vec{\textbf{r}} = (x,y)^\intercal$. In our resonator these are partitioned into three separately degenerate sets corresponding to lowest Landau levels on different cones. These sets are the $l  = 0, 3, 6, \ldots$ modes, the $l = 1, 4, 7, \ldots$ modes, and the $l = 2, 5, 8, \ldots$ modes and satisfy the angular symmetry condition $\psi_l(\theta + \frac{2\pi}{3}) = e^{2\pi i c/3}\psi_l(\theta)$, where $c = 0, 1$, or $2$ is the lowest angular momentum state in the set and serves as the cone's label. c = 0 defines the symmetry relation that describes an unthreaded cone; with $c \neq 0$, the cone has an additional Aharanov-Bohm phase arising from $c/3$ magnetic flux quanta threaded through its tip (Fig. 3c). Angular momentum states encircling the cone tip enclose this flux three times, so states experience integer flux, reflected in their $\sqrt{l}$ radial extension.

Away from the apex, photons on each cone behave as in a flat space lowest Landau level. In Fig. 3d, we identify each cone by the lowest angular momentum state supported around its apex. Then, on each cone, we show that we can create arbitrary angular momentum states ($l = 0, 1$) about displaced points so long as the displaced mode does not self intersect or encircle the cone tip. Beyond reflecting the invariance of our system under magnetic translations, this permits the creation of canonical fractional quantum Hall states in a future interacting system, in addition to novel Laughlin states accessible at the cone tip (see Supplementary Information). As a visualization, figures 3e-f project these displaced $l = 0$ and $l = 1$ modes onto a cone, further demonstrating that, away from the apex, modes on the cone closely resemble modes on a regular plane. 

The topological numbers that characterize quantum Hall phases are predicted to specify the response of the photonic local density of states (LDOS) to magnetic field and spatial curvature, as described by the Wen-Zee theory\cite{Can2014, Wen1992, Hoyos2012, Abanov2014} (see Supplementary Information). We perform an experimental test of this theory by measuring the LDOS (Figs. 3g-i) via transmission images of each state in the relevant weakly split Landau level and summing these images (see Supplementary Information). We then compare the LDOS near the cone tip with the flat space density away from the tip (within each panel Fig. 3g-i) and compare the LDOS with different quantities of flux threaded (between panels Fig. 3g-i). We clearly observe a density buildup for the $c = 0$ cone; however, we find a vanishing LDOS on the other two cones, reflecting additional magnetic flux threaded through their tips equal to $-\Phi_0/3$ and $-2\Phi_0/3$, where $\Phi_0$ is the magnetic flux quantum (Fig. 3c). According to the Wen-Zee effective theory, the expected excess state number is given by $\delta N = \frac{2}{3} \bar{s} - \frac{c}{3}$, where $c/3$ is the number of flux quanta threaded through the cone tip and $\bar{s}$ is a parameter called the mean orbital spin that characterizes particles’ coupling to spatial curvature and is predicted to be $1/2$ for the lowest Landau level\cite{Wen1992} (see Supplementary Information). We therefore expect $\delta N = 1/3, 0$, and $-1/3$ of a state near the tips of the $c = 0, 1,$ and $2$ cones, respectively. By integrating the measured LDOS excess or deficit near the apex, we measure the state number excess to be $0.31(2)$ on the $c = 0$ cone, $-0.02(1)$ on the $c = 1$ cone, and $-0.35(2)$ on the $c = 2$ cone, yielding the experimentally measured value $\bar{s} = 0.47(1)$. We find quantitative agreement between our measured results and the Wen-Zee  theory. 

We have demonstrated a synthetic magnetic field for continuum photons. Furthermore, we have created an integer quantum Hall system in curved space, a long-standing challenge in condensed matter physics. We can extend our tests of the Wen-Zee theory by measuring fractional state number excess in higher Landau levels and examining the connection between the mean orbital spin and the Hall viscosity\cite{Read2009} (see Supplementary Information). Our approach clears a path to the photonic fractional quantum Hall regime, as it is compatible with Rydberg-mediated strong photon-photon interactions\cite{Sommer2015}, and does not require the low particle densities (and thus weakened interactions) necessary to map Laughlin physics onto a lattice. Simply avoiding the cone apex will allow the spectroscopic creation and detection of flat space fractional quantum hall states such as the Laughlin wavefunction (see Supplementary Information), while exploring the apex will afford the opportunity to investigate the interplay of geometry and topology in strongly correlated quantum materials.

We acknowledge conversations with I. Carusotto, M. Levin, and P. Wiegmann. This work was supported by DOE, DARPA, and AFOSR. A.G. acknowledges the support of the Kadanoff Center for Theoretical Physics. A.R. acknowledges support from ARO through an NDSEG fellowship. The experiment was designed and built by N.S., J.S., A.R., and A.S. Measurement and analysis of the data was performed by N.S. Theoretical development and interpretation of results were performed by J.S., A.S., N.S., and A.G. All authors contributed to the manuscript. Correspondence and requests for materials should be addressed to J.S. (simonjon@uchicago.edu).



\bibliography{library.bib}

\begin{thebibliography}{30}%
\makeatletter
\providecommand \@ifxundefined [1]{%
 \@ifx{#1\undefined}
}%
\providecommand \@ifnum [1]{%
 \ifnum #1\expandafter \@firstoftwo
 \else \expandafter \@secondoftwo
 \fi
}%
\providecommand \@ifx [1]{%
 \ifx #1\expandafter \@firstoftwo
 \else \expandafter \@secondoftwo
 \fi
}%
\providecommand \natexlab [1]{#1}%
\providecommand \enquote  [1]{``#1''}%
\providecommand \bibnamefont  [1]{#1}%
\providecommand \bibfnamefont [1]{#1}%
\providecommand \citenamefont [1]{#1}%
\providecommand \href@noop [0]{\@secondoftwo}%
\providecommand \href [0]{\begingroup \@sanitize@url \@href}%
\providecommand \@href[1]{\@@startlink{#1}\@@href}%
\providecommand \@@href[1]{\endgroup#1\@@endlink}%
\providecommand \@sanitize@url [0]{\catcode `\\12\catcode `\$12\catcode
  `\&12\catcode `\#12\catcode `\^12\catcode `\_12\catcode `\%12\relax}%
\providecommand \@@startlink[1]{}%
\providecommand \@@endlink[0]{}%
\providecommand \url  [0]{\begingroup\@sanitize@url \@url }%
\providecommand \@url [1]{\endgroup\@href {#1}{\urlprefix }}%
\providecommand \urlprefix  [0]{URL }%
\providecommand \Eprint [0]{\href }%
\providecommand \doibase [0]{http://dx.doi.org/}%
\providecommand \selectlanguage [0]{\@gobble}%
\providecommand \bibinfo  [0]{\@secondoftwo}%
\providecommand \bibfield  [0]{\@secondoftwo}%
\providecommand \translation [1]{[#1]}%
\providecommand \BibitemOpen [0]{}%
\providecommand \bibitemStop [0]{}%
\providecommand \bibitemNoStop [0]{.\EOS\space}%
\providecommand \EOS [0]{\spacefactor3000\relax}%
\providecommand \BibitemShut  [1]{\csname bibitem#1\endcsname}%
\let\auto@bib@innerbib\@empty
\bibitem [{\citenamefont {Peyronel}\ \emph {et~al.}(2012)\citenamefont
  {Peyronel}, \citenamefont {Firstenberg}, \citenamefont {Liang}, \citenamefont
  {Hofferberth}, \citenamefont {Gorshkov}, \citenamefont {Pohl}, \citenamefont
  {Lukin},\ and\ \citenamefont {Vuletić}}]{Peyronel2012}%
  \BibitemOpen
  \bibfield  {author} {\bibinfo {author} {\bibfnamefont {T.}~\bibnamefont
  {Peyronel}}, \bibinfo {author} {\bibfnamefont {O.}~\bibnamefont
  {Firstenberg}}, \bibinfo {author} {\bibfnamefont {Q.-Y.}\ \bibnamefont
  {Liang}}, \bibinfo {author} {\bibfnamefont {S.}~\bibnamefont {Hofferberth}},
  \bibinfo {author} {\bibfnamefont {A.~V.}\ \bibnamefont {Gorshkov}}, \bibinfo
  {author} {\bibfnamefont {T.}~\bibnamefont {Pohl}}, \bibinfo {author}
  {\bibfnamefont {M.~D.}\ \bibnamefont {Lukin}}, \ and\ \bibinfo {author}
  {\bibfnamefont {V.}~\bibnamefont {Vuletić}},\ }\href {\doibase
  10.1038/nature11361} {\bibfield  {journal} {\bibinfo  {journal} {Nature}\
  }\textbf {\bibinfo {volume} {488}},\ \bibinfo {pages} {57} (\bibinfo {year}
  {2012})}\BibitemShut {NoStop}%
\bibitem [{\citenamefont {Gopalakrishnan}\ \emph {et~al.}(2009)\citenamefont
  {Gopalakrishnan}, \citenamefont {Lev},\ and\ \citenamefont
  {Goldbart}}]{Gopalakrishnan2009}%
  \BibitemOpen
  \bibfield  {author} {\bibinfo {author} {\bibfnamefont {S.}~\bibnamefont
  {Gopalakrishnan}}, \bibinfo {author} {\bibfnamefont {B.~L.}\ \bibnamefont
  {Lev}}, \ and\ \bibinfo {author} {\bibfnamefont {P.~M.}\ \bibnamefont
  {Goldbart}},\ }\href {\doibase 10.1038/nphys1403} {\bibfield  {journal}
  {\bibinfo  {journal} {Nature Physics}\ }\textbf {\bibinfo {volume} {5}},\
  \bibinfo {pages} {845} (\bibinfo {year} {2009})}\BibitemShut {NoStop}%
\bibitem [{\citenamefont {Baumann}\ \emph {et~al.}(2010)\citenamefont
  {Baumann}, \citenamefont {Guerlin}, \citenamefont {Brennecke},\ and\
  \citenamefont {Esslinger}}]{Baumann2010}%
  \BibitemOpen
  \bibfield  {author} {\bibinfo {author} {\bibfnamefont {K.}~\bibnamefont
  {Baumann}}, \bibinfo {author} {\bibfnamefont {C.}~\bibnamefont {Guerlin}},
  \bibinfo {author} {\bibfnamefont {F.}~\bibnamefont {Brennecke}}, \ and\
  \bibinfo {author} {\bibfnamefont {T.}~\bibnamefont {Esslinger}},\ }\href
  {\doibase 10.1038/nature09009} {\bibfield  {journal} {\bibinfo  {journal}
  {Nature}\ }\textbf {\bibinfo {volume} {464}},\ \bibinfo {pages} {1301}
  (\bibinfo {year} {2010})}\BibitemShut {NoStop}%
\bibitem [{\citenamefont {Cooper}\ and\ \citenamefont
  {Dalibard}(2013)}]{Cooper2013}%
  \BibitemOpen
  \bibfield  {author} {\bibinfo {author} {\bibfnamefont {N.~R.}\ \bibnamefont
  {Cooper}}\ and\ \bibinfo {author} {\bibfnamefont {J.}~\bibnamefont
  {Dalibard}},\ }\href {http://arxiv.org/abs/1212.3552} {\bibfield  {journal}
  {\bibinfo  {journal} {Physical Review Letters}\ }\textbf {\bibinfo {volume}
  {110}},\ \bibinfo {pages} {185301} (\bibinfo {year} {2013})}\BibitemShut
  {NoStop}%
\bibitem [{\citenamefont {Carusotto}\ and\ \citenamefont
  {Ciuti}(2013)}]{Carusotto2013}%
  \BibitemOpen
  \bibfield  {author} {\bibinfo {author} {\bibfnamefont {I.}~\bibnamefont
  {Carusotto}}\ and\ \bibinfo {author} {\bibfnamefont {C.}~\bibnamefont
  {Ciuti}},\ }\href {\doibase 10.1103/RevModPhys.85.299} {\bibfield  {journal}
  {\bibinfo  {journal} {Reviews of Modern Physics}\ }\textbf {\bibinfo {volume}
  {85}},\ \bibinfo {pages} {299} (\bibinfo {year} {2013})}\BibitemShut
  {NoStop}%
\bibitem [{\citenamefont {Jia}\ \emph {et~al.}(2015)\citenamefont {Jia},
  \citenamefont {Owens}, \citenamefont {Sommer}, \citenamefont {Schuster},\
  and\ \citenamefont {Simon}}]{Jia2015b}%
  \BibitemOpen
  \bibfield  {author} {\bibinfo {author} {\bibfnamefont {N.}~\bibnamefont
  {Jia}}, \bibinfo {author} {\bibfnamefont {C.}~\bibnamefont {Owens}}, \bibinfo
  {author} {\bibfnamefont {A.}~\bibnamefont {Sommer}}, \bibinfo {author}
  {\bibfnamefont {D.}~\bibnamefont {Schuster}}, \ and\ \bibinfo {author}
  {\bibfnamefont {J.}~\bibnamefont {Simon}},\ }\href
  {http://arxiv.org/abs/1309.0878} {\bibfield  {journal} {\bibinfo  {journal}
  {Phys. Rev. X}\ }\textbf {\bibinfo {volume} {5}} (\bibinfo {year}
  {2015})}\BibitemShut {NoStop}%
\bibitem [{\citenamefont {Otterbach}\ \emph {et~al.}(2010)\citenamefont
  {Otterbach}, \citenamefont {Ruseckas}, \citenamefont {Unanyan}, \citenamefont
  {Juzeliūnas},\ and\ \citenamefont {Fleischhauer}}]{Otterbach2010}%
  \BibitemOpen
  \bibfield  {author} {\bibinfo {author} {\bibfnamefont {J.}~\bibnamefont
  {Otterbach}}, \bibinfo {author} {\bibfnamefont {J.}~\bibnamefont {Ruseckas}},
  \bibinfo {author} {\bibfnamefont {R.~G.}\ \bibnamefont {Unanyan}}, \bibinfo
  {author} {\bibfnamefont {G.}~\bibnamefont {Juzeliūnas}}, \ and\ \bibinfo
  {author} {\bibfnamefont {M.}~\bibnamefont {Fleischhauer}},\ }\href
  {http://link.aps.org/doi/10.1103/PhysRevLett.104.033903
  http://journals.aps.org/prl/abstract/10.1103/PhysRevLett.104.033903}
  {\bibfield  {journal} {\bibinfo  {journal} {Physical Review Letters}\
  }\textbf {\bibinfo {volume} {104}},\ \bibinfo {pages} {033903} (\bibinfo
  {year} {2010})}\BibitemShut {NoStop}%
\bibitem [{\citenamefont {Wang}\ \emph {et~al.}(2009)\citenamefont {Wang},
  \citenamefont {Chong}, \citenamefont {Joannopoulos},\ and\ \citenamefont
  {Solja{\u c}i{\' c}}}]{Wang2009}%
  \BibitemOpen
  \bibfield  {author} {\bibinfo {author} {\bibfnamefont {Z.}~\bibnamefont
  {Wang}}, \bibinfo {author} {\bibfnamefont {Y.}~\bibnamefont {Chong}},
  \bibinfo {author} {\bibfnamefont {J.~D.}\ \bibnamefont {Joannopoulos}}, \
  and\ \bibinfo {author} {\bibfnamefont {M.}~\bibnamefont {Solja{\u c}i{\'
  c}}},\ }\href {\doibase
  http://www.nature.com/nature/journal/v461/n7265/suppinfo/nature08293_S1.html}
  {\bibfield  {journal} {\bibinfo  {journal} {Nature}\ }\textbf {\bibinfo
  {volume} {461}},\ \bibinfo {pages} {772} (\bibinfo {year}
  {2009})}\BibitemShut {NoStop}%
\bibitem [{\citenamefont {Yuan}\ \emph {et~al.}(2007)\citenamefont {Yuan},
  \citenamefont {Long}, \citenamefont {Liang}, \citenamefont {Zhang},
  \citenamefont {Wang},\ and\ \citenamefont {Zhao}}]{Yuan2007}%
  \BibitemOpen
  \bibfield  {author} {\bibinfo {author} {\bibfnamefont {J.}~\bibnamefont
  {Yuan}}, \bibinfo {author} {\bibfnamefont {X.}~\bibnamefont {Long}}, \bibinfo
  {author} {\bibfnamefont {L.}~\bibnamefont {Liang}}, \bibinfo {author}
  {\bibfnamefont {B.}~\bibnamefont {Zhang}}, \bibinfo {author} {\bibfnamefont
  {F.}~\bibnamefont {Wang}}, \ and\ \bibinfo {author} {\bibfnamefont
  {H.}~\bibnamefont {Zhao}},\ }\href {\doibase 10.1364/AO.46.002980} {\bibfield
   {journal} {\bibinfo  {journal} {Applied Optics}\ }\textbf {\bibinfo {volume}
  {46}},\ \bibinfo {pages} {2980} (\bibinfo {year} {2007})}\BibitemShut
  {NoStop}%
\bibitem [{\citenamefont {Cooper}(2008)}]{Cooper2008}%
  \BibitemOpen
  \bibfield  {author} {\bibinfo {author} {\bibfnamefont {N.~R.}\ \bibnamefont
  {Cooper}},\ }\href {\doibase 10.1080/00018730802564122} {\bibfield  {journal}
  {\bibinfo  {journal} {Advances in Physics}\ }\textbf {\bibinfo {volume}
  {57}},\ \bibinfo {pages} {539} (\bibinfo {year} {2008})}\BibitemShut
  {NoStop}%
\bibitem [{\citenamefont {Bloch}\ \emph {et~al.}(2008)\citenamefont {Bloch},
  \citenamefont {Dalibard},\ and\ \citenamefont {Zwerger}}]{Bloch2008}%
  \BibitemOpen
  \bibfield  {author} {\bibinfo {author} {\bibfnamefont {I.}~\bibnamefont
  {Bloch}}, \bibinfo {author} {\bibfnamefont {J.}~\bibnamefont {Dalibard}}, \
  and\ \bibinfo {author} {\bibfnamefont {W.}~\bibnamefont {Zwerger}},\ }\href
  {http://link.aps.org/doi/10.1103/RevModPhys.80.885} {\bibfield  {journal}
  {\bibinfo  {journal} {Reviews of Modern Physics}\ }\textbf {\bibinfo {volume}
  {80}},\ \bibinfo {pages} {885} (\bibinfo {year} {2008})}\BibitemShut
  {NoStop}%
\bibitem [{\citenamefont {Wen}\ and\ \citenamefont {Zee}(1992)}]{Wen1992}%
  \BibitemOpen
  \bibfield  {author} {\bibinfo {author} {\bibfnamefont {X.~G.}\ \bibnamefont
  {Wen}}\ and\ \bibinfo {author} {\bibfnamefont {A.}~\bibnamefont {Zee}},\
  }\href {http://link.aps.org/doi/10.1103/PhysRevLett.69.953} {\bibfield
  {journal} {\bibinfo  {journal} {Physical Review Letters}\ }\textbf {\bibinfo
  {volume} {69}},\ \bibinfo {pages} {953} (\bibinfo {year} {1992})}\BibitemShut
  {NoStop}%
\bibitem [{\citenamefont {Hoyos}\ and\ \citenamefont {Son}(2012)}]{Hoyos2012}%
  \BibitemOpen
  \bibfield  {author} {\bibinfo {author} {\bibfnamefont {C.}~\bibnamefont
  {Hoyos}}\ and\ \bibinfo {author} {\bibfnamefont {D.~T.}\ \bibnamefont
  {Son}},\ }\href {http://link.aps.org/doi/10.1103/PhysRevLett.108.066805}
  {\bibfield  {journal} {\bibinfo  {journal} {Physical Review Letters}\
  }\textbf {\bibinfo {volume} {108}},\ \bibinfo {pages} {066805} (\bibinfo
  {year} {2012})}\BibitemShut {NoStop}%
\bibitem [{\citenamefont {Abanov}\ and\ \citenamefont
  {Gromov}(2014)}]{Abanov2014}%
  \BibitemOpen
  \bibfield  {author} {\bibinfo {author} {\bibfnamefont {A.~G.}\ \bibnamefont
  {Abanov}}\ and\ \bibinfo {author} {\bibfnamefont {A.}~\bibnamefont
  {Gromov}},\ }\href {http://link.aps.org/doi/10.1103/PhysRevB.90.014435}
  {\bibfield  {journal} {\bibinfo  {journal} {Physical Review B}\ }\textbf
  {\bibinfo {volume} {90}},\ \bibinfo {pages} {014435} (\bibinfo {year}
  {2014})}\BibitemShut {NoStop}%
\bibitem [{\citenamefont {Can}\ \emph {et~al.}(2014)\citenamefont {Can},
  \citenamefont {Laskin},\ and\ \citenamefont {Wiegmann}}]{Can2014}%
  \BibitemOpen
  \bibfield  {author} {\bibinfo {author} {\bibfnamefont {T.}~\bibnamefont
  {Can}}, \bibinfo {author} {\bibfnamefont {M.}~\bibnamefont {Laskin}}, \ and\
  \bibinfo {author} {\bibfnamefont {P.}~\bibnamefont {Wiegmann}},\ }\href@noop
  {} {\bibfield  {journal} {\bibinfo  {journal} {Physical Review Letters}\
  }\textbf {\bibinfo {volume} {113}},\ \bibinfo {pages} {046803} (\bibinfo
  {year} {2014})}\BibitemShut {NoStop}%
\bibitem [{\citenamefont {Sommer}\ \emph {et~al.}(2015)\citenamefont {Sommer},
  \citenamefont {B{\"u}chler},\ and\ \citenamefont {Simon}}]{Sommer2015}%
  \BibitemOpen
  \bibfield  {author} {\bibinfo {author} {\bibfnamefont {A.}~\bibnamefont
  {Sommer}}, \bibinfo {author} {\bibfnamefont {H.~P.}\ \bibnamefont
  {B{\"u}chler}}, \ and\ \bibinfo {author} {\bibfnamefont {J.}~\bibnamefont
  {Simon}},\ }\href@noop {} {\bibfield  {journal} {\bibinfo  {journal} {arXiv
  preprint arXiv:1506.00341}\ } (\bibinfo {year} {2015})}\BibitemShut {NoStop}%
\bibitem [{\citenamefont {Umucal{\i}lar}\ \emph {et~al.}(2014)\citenamefont
  {Umucal{\i}lar}, \citenamefont {Wouters},\ and\ \citenamefont
  {Carusotto}}]{Umucalilar2014}%
  \BibitemOpen
  \bibfield  {author} {\bibinfo {author} {\bibfnamefont {R.~O.}\ \bibnamefont
  {Umucal{\i}lar}}, \bibinfo {author} {\bibfnamefont {M.}~\bibnamefont
  {Wouters}}, \ and\ \bibinfo {author} {\bibfnamefont {I.}~\bibnamefont
  {Carusotto}},\ }\href {http://link.aps.org/doi/10.1103/PhysRevA.89.023803}
  {\bibfield  {journal} {\bibinfo  {journal} {Physical Review A}\ }\textbf
  {\bibinfo {volume} {89}},\ \bibinfo {pages} {023803} (\bibinfo {year}
  {2014})}\BibitemShut {NoStop}%
\bibitem [{\citenamefont {Paredes}\ \emph {et~al.}(2001)\citenamefont
  {Paredes}, \citenamefont {Fedichev}, \citenamefont {Cirac},\ and\
  \citenamefont {Zoller}}]{Paredes2001}%
  \BibitemOpen
  \bibfield  {author} {\bibinfo {author} {\bibfnamefont {B.}~\bibnamefont
  {Paredes}}, \bibinfo {author} {\bibfnamefont {P.}~\bibnamefont {Fedichev}},
  \bibinfo {author} {\bibfnamefont {J.~I.}\ \bibnamefont {Cirac}}, \ and\
  \bibinfo {author} {\bibfnamefont {P.}~\bibnamefont {Zoller}},\ }\href
  {http://link.aps.org/doi/10.1103/PhysRevLett.87.010402
  http://journals.aps.org/prl/abstract/10.1103/PhysRevLett.87.010402}
  {\bibfield  {journal} {\bibinfo  {journal} {Physical Review Letters}\
  }\textbf {\bibinfo {volume} {87}},\ \bibinfo {pages} {010402} (\bibinfo
  {year} {2001})}\BibitemShut {NoStop}%
\bibitem [{\citenamefont {Umucal{\i}lar}\ and\ \citenamefont
  {Carusotto}(2013)}]{Umucalilar2013}%
  \BibitemOpen
  \bibfield  {author} {\bibinfo {author} {\bibfnamefont {R.~O.}\ \bibnamefont
  {Umucal{\i}lar}}\ and\ \bibinfo {author} {\bibfnamefont {I.}~\bibnamefont
  {Carusotto}},\ }\href {\doibase
  http://dx.doi.org/10.1016/j.physleta.2013.06.011} {\bibfield  {journal}
  {\bibinfo  {journal} {Physics Letters A}\ }\textbf {\bibinfo {volume}
  {377}},\ \bibinfo {pages} {2074} (\bibinfo {year} {2013})}\BibitemShut
  {NoStop}%
\bibitem [{\citenamefont {Nayak}\ \emph {et~al.}(2008)\citenamefont {Nayak},
  \citenamefont {Simon}, \citenamefont {Stern}, \citenamefont {Freedman},\ and\
  \citenamefont {Das~Sarma}}]{Nayak2008}%
  \BibitemOpen
  \bibfield  {author} {\bibinfo {author} {\bibfnamefont {C.}~\bibnamefont
  {Nayak}}, \bibinfo {author} {\bibfnamefont {S.~H.}\ \bibnamefont {Simon}},
  \bibinfo {author} {\bibfnamefont {A.}~\bibnamefont {Stern}}, \bibinfo
  {author} {\bibfnamefont {M.}~\bibnamefont {Freedman}}, \ and\ \bibinfo
  {author} {\bibfnamefont {S.}~\bibnamefont {Das~Sarma}},\ }\href
  {http://link.aps.org/doi/10.1103/RevModPhys.80.1083} {\bibfield  {journal}
  {\bibinfo  {journal} {Reviews of Modern Physics}\ }\textbf {\bibinfo {volume}
  {80}},\ \bibinfo {pages} {1083} (\bibinfo {year} {2008})}\BibitemShut
  {NoStop}%
\bibitem [{\citenamefont {Wang}\ \emph {et~al.}(2008)\citenamefont {Wang},
  \citenamefont {Chong}, \citenamefont {Joannopoulos},\ and\ \citenamefont
  {Solja{\u c}i{\' c}}}]{Wang2008}%
  \BibitemOpen
  \bibfield  {author} {\bibinfo {author} {\bibfnamefont {Z.}~\bibnamefont
  {Wang}}, \bibinfo {author} {\bibfnamefont {Y.~D.}\ \bibnamefont {Chong}},
  \bibinfo {author} {\bibfnamefont {J.~D.}\ \bibnamefont {Joannopoulos}}, \
  and\ \bibinfo {author} {\bibfnamefont {M.}~\bibnamefont {Solja{\u c}i{\'
  c}}},\ }\href {http://link.aps.org/doi/10.1103/PhysRevLett.100.013905
  http://journals.aps.org/prl/abstract/10.1103/PhysRevLett.100.013905}
  {\bibfield  {journal} {\bibinfo  {journal} {Physical Review Letters}\
  }\textbf {\bibinfo {volume} {100}},\ \bibinfo {pages} {013905} (\bibinfo
  {year} {2008})}\BibitemShut {NoStop}%
\bibitem [{\citenamefont {Rechtsman}\ \emph
  {et~al.}(2013{\natexlab{a}})\citenamefont {Rechtsman}, \citenamefont
  {Zeuner}, \citenamefont {Plotnik}, \citenamefont {Lumer}, \citenamefont
  {Podolsky}, \citenamefont {Dreisow}, \citenamefont {Nolte}, \citenamefont
  {Segev},\ and\ \citenamefont {Szameit}}]{Rechtsman2013}%
  \BibitemOpen
  \bibfield  {author} {\bibinfo {author} {\bibfnamefont {M.~C.}\ \bibnamefont
  {Rechtsman}}, \bibinfo {author} {\bibfnamefont {J.~M.}\ \bibnamefont
  {Zeuner}}, \bibinfo {author} {\bibfnamefont {Y.}~\bibnamefont {Plotnik}},
  \bibinfo {author} {\bibfnamefont {Y.}~\bibnamefont {Lumer}}, \bibinfo
  {author} {\bibfnamefont {D.}~\bibnamefont {Podolsky}}, \bibinfo {author}
  {\bibfnamefont {F.}~\bibnamefont {Dreisow}}, \bibinfo {author} {\bibfnamefont
  {S.}~\bibnamefont {Nolte}}, \bibinfo {author} {\bibfnamefont
  {M.}~\bibnamefont {Segev}}, \ and\ \bibinfo {author} {\bibfnamefont
  {A.}~\bibnamefont {Szameit}},\ }\href {\doibase 10.1038/nature12066}
  {\bibfield  {journal} {\bibinfo  {journal} {Nature}\ }\textbf {\bibinfo
  {volume} {496}},\ \bibinfo {pages} {196} (\bibinfo {year}
  {2013}{\natexlab{a}})}\BibitemShut {NoStop}%
\bibitem [{\citenamefont {Hafezi}\ \emph {et~al.}(2013)\citenamefont {Hafezi},
  \citenamefont {Lukin},\ and\ \citenamefont {Taylor}}]{Hafezi2013}%
  \BibitemOpen
  \bibfield  {author} {\bibinfo {author} {\bibfnamefont {M.}~\bibnamefont
  {Hafezi}}, \bibinfo {author} {\bibfnamefont {M.~D.}\ \bibnamefont {Lukin}}, \
  and\ \bibinfo {author} {\bibfnamefont {J.~M.}\ \bibnamefont {Taylor}},\
  }\href {http://iopscience.iop.org/1367-2630/15/6/063001
  http://iopscience.iop.org/1367-2630/15/6/063001/pdf/1367-2630_15_6_063001.pdf}
  {\bibfield  {journal} {\bibinfo  {journal} {New Journal of Physics}\ }\textbf
  {\bibinfo {volume} {15}},\ \bibinfo {pages} {063001} (\bibinfo {year}
  {2013})}\BibitemShut {NoStop}%
\bibitem [{\citenamefont {Rechtsman}\ \emph
  {et~al.}(2013{\natexlab{b}})\citenamefont {Rechtsman}, \citenamefont
  {Zeuner}, \citenamefont {Plotnik}, \citenamefont {Lumer}, \citenamefont
  {Podolsky}, \citenamefont {Dreisow}, \citenamefont {Nolte}, \citenamefont
  {Segev},\ and\ \citenamefont {Szameit}}]{Rechtsman2013b}%
  \BibitemOpen
  \bibfield  {author} {\bibinfo {author} {\bibfnamefont {M.~C.}\ \bibnamefont
  {Rechtsman}}, \bibinfo {author} {\bibfnamefont {J.~M.}\ \bibnamefont
  {Zeuner}}, \bibinfo {author} {\bibfnamefont {Y.}~\bibnamefont {Plotnik}},
  \bibinfo {author} {\bibfnamefont {Y.}~\bibnamefont {Lumer}}, \bibinfo
  {author} {\bibfnamefont {D.}~\bibnamefont {Podolsky}}, \bibinfo {author}
  {\bibfnamefont {F.}~\bibnamefont {Dreisow}}, \bibinfo {author} {\bibfnamefont
  {S.}~\bibnamefont {Nolte}}, \bibinfo {author} {\bibfnamefont
  {M.}~\bibnamefont {Segev}}, \ and\ \bibinfo {author} {\bibfnamefont
  {A.}~\bibnamefont {Szameit}},\ }\href {\doibase 10.1038/nature12066}
  {\bibfield  {journal} {\bibinfo  {journal} {Nature}\ }\textbf {\bibinfo
  {volume} {496}},\ \bibinfo {pages} {196} (\bibinfo {year}
  {2013}{\natexlab{b}})}\BibitemShut {NoStop}%
\bibitem [{\citenamefont {Karzig}\ \emph {et~al.}(2015)\citenamefont {Karzig},
  \citenamefont {Bardyn}, \citenamefont {Lindner},\ and\ \citenamefont
  {Refael}}]{Karzig2015}%
  \BibitemOpen
  \bibfield  {author} {\bibinfo {author} {\bibfnamefont {T.}~\bibnamefont
  {Karzig}}, \bibinfo {author} {\bibfnamefont {C.-E.}\ \bibnamefont {Bardyn}},
  \bibinfo {author} {\bibfnamefont {N.~H.}\ \bibnamefont {Lindner}}, \ and\
  \bibinfo {author} {\bibfnamefont {G.}~\bibnamefont {Refael}},\ }\href
  {http://link.aps.org/doi/10.1103/PhysRevX.5.031001} {\bibfield  {journal}
  {\bibinfo  {journal} {Physical Review X}\ }\textbf {\bibinfo {volume} {5}},\
  \bibinfo {pages} {031001} (\bibinfo {year} {2015})}\BibitemShut {NoStop}%
\bibitem [{\citenamefont {Longhi}(2015)}]{Longhi2015}%
  \BibitemOpen
  \bibfield  {author} {\bibinfo {author} {\bibfnamefont {S.}~\bibnamefont
  {Longhi}},\ }\href {\doibase 10.1364/OL.40.002941} {\bibfield  {journal}
  {\bibinfo  {journal} {Optics Letters}\ }\textbf {\bibinfo {volume} {40}},\
  \bibinfo {pages} {2941} (\bibinfo {year} {2015})}\BibitemShut {NoStop}%
\bibitem [{\citenamefont {Klaers}\ \emph {et~al.}(2010)\citenamefont {Klaers},
  \citenamefont {Schmitt}, \citenamefont {Vewinger},\ and\ \citenamefont
  {Weitz}}]{Klaers2010}%
  \BibitemOpen
  \bibfield  {author} {\bibinfo {author} {\bibfnamefont {J.}~\bibnamefont
  {Klaers}}, \bibinfo {author} {\bibfnamefont {J.}~\bibnamefont {Schmitt}},
  \bibinfo {author} {\bibfnamefont {F.}~\bibnamefont {Vewinger}}, \ and\
  \bibinfo {author} {\bibfnamefont {M.}~\bibnamefont {Weitz}},\ }\href
  {\doibase 10.1038/nature09567} {\bibfield  {journal} {\bibinfo  {journal}
  {Nature}\ }\textbf {\bibinfo {volume} {468}},\ \bibinfo {pages} {545}
  (\bibinfo {year} {2010})}\BibitemShut {NoStop}%
\bibitem [{\citenamefont {Sommer}\ and\ \citenamefont
  {Simon}(2016)}]{Sommer2016}%
  \BibitemOpen
  \bibfield  {author} {\bibinfo {author} {\bibfnamefont {A.}~\bibnamefont
  {Sommer}}\ and\ \bibinfo {author} {\bibfnamefont {J.}~\bibnamefont {Simon}},\
  }\href {http://stacks.iop.org/1367-2630/18/i=3/a=035008} {\bibfield
  {journal} {\bibinfo  {journal} {New Journal of Physics}\ }\textbf {\bibinfo
  {volume} {18}},\ \bibinfo {pages} {035008} (\bibinfo {year}
  {2016})}\BibitemShut {NoStop}%
\bibitem [{\citenamefont {Schweikhard}\ \emph {et~al.}(2004)\citenamefont
  {Schweikhard}, \citenamefont {Coddington}, \citenamefont {Engels},
  \citenamefont {Mogendorff},\ and\ \citenamefont {Cornell}}]{Schweikhard2004}%
  \BibitemOpen
  \bibfield  {author} {\bibinfo {author} {\bibfnamefont {V.}~\bibnamefont
  {Schweikhard}}, \bibinfo {author} {\bibfnamefont {I.}~\bibnamefont
  {Coddington}}, \bibinfo {author} {\bibfnamefont {P.}~\bibnamefont {Engels}},
  \bibinfo {author} {\bibfnamefont {V.~P.}\ \bibnamefont {Mogendorff}}, \ and\
  \bibinfo {author} {\bibfnamefont {E.~A.}\ \bibnamefont {Cornell}},\ }\href
  {http://link.aps.org/doi/10.1103/PhysRevLett.92.040404
  http://journals.aps.org/prl/abstract/10.1103/PhysRevLett.92.040404}
  {\bibfield  {journal} {\bibinfo  {journal} {Physical Review Letters}\
  }\textbf {\bibinfo {volume} {92}},\ \bibinfo {pages} {040404} (\bibinfo
  {year} {2004})}\BibitemShut {NoStop}%
\bibitem [{\citenamefont {Read}(2009)}]{Read2009}%
  \BibitemOpen
  \bibfield  {author} {\bibinfo {author} {\bibfnamefont {N.}~\bibnamefont
  {Read}},\ }\href {http://link.aps.org/doi/10.1103/PhysRevB.79.045308}
  {\bibfield  {journal} {\bibinfo  {journal} {Physical Review B}\ }\textbf
  {\bibinfo {volume} {79}},\ \bibinfo {pages} {045308} (\bibinfo {year}
  {2009})}\BibitemShut {NoStop}%
\end{thebibliography}%



\newpage
\clearpage
\newpage

\onecolumngrid

\includegraphics[page=1, trim=70 70 70 70,clip,width=1\linewidth]{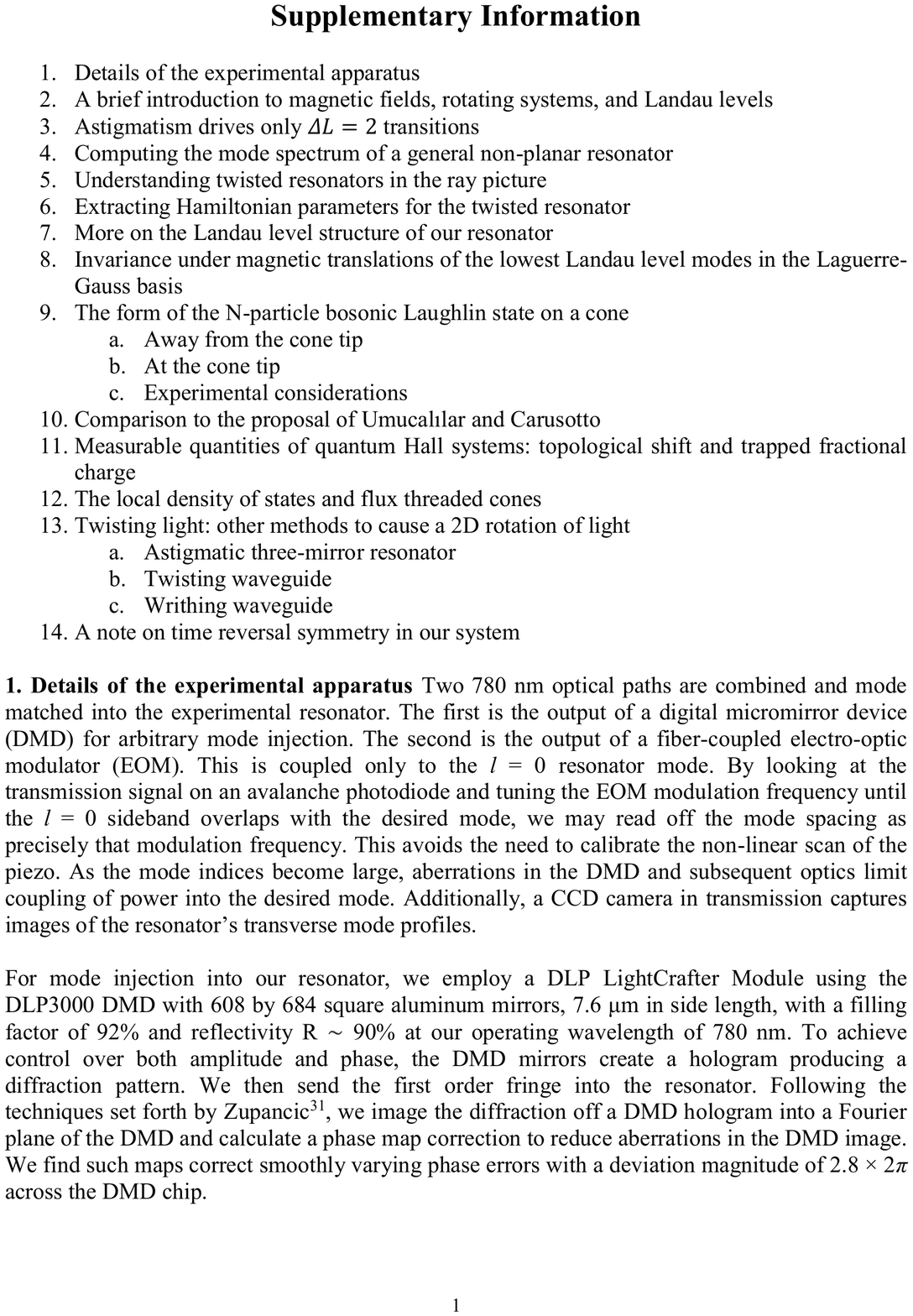}
\includegraphics[page=2, trim=70 70 70 70,clip,width=1\linewidth]{Schine-suppinfo_Revised_3.pdf}
\includegraphics[page=3, trim=70 70 70 70,clip,width=1\linewidth]{Schine-suppinfo_Revised_3.pdf}
\includegraphics[page=4, trim=70 70 70 70,clip,width=1\linewidth]{Schine-suppinfo_Revised_3.pdf}
\includegraphics[page=5, trim=70 70 70 70,clip,width=1\linewidth]{Schine-suppinfo_Revised_3.pdf}
\includegraphics[page=6, trim=70 70 70 70,clip,width=1\linewidth]{Schine-suppinfo_Revised_3.pdf}
\includegraphics[page=7, trim=70 70 70 70,clip,width=1\linewidth]{Schine-suppinfo_Revised_3.pdf}
\includegraphics[page=8, trim=70 70 70 70,clip,width=1\linewidth]{Schine-suppinfo_Revised_3.pdf}
\includegraphics[page=9, trim=70 70 70 70,clip,width=1\linewidth]{Schine-suppinfo_Revised_3.pdf}
\includegraphics[page=10, trim=70 70 70 70,clip,width=1\linewidth]{Schine-suppinfo_Revised_3.pdf}
\includegraphics[page=11, trim=70 70 70 70,clip,width=1\linewidth]{Schine-suppinfo_Revised_3.pdf}
\includegraphics[page=12, trim=70 70 70 70,clip,width=1\linewidth]{Schine-suppinfo_Revised_3.pdf}
\includegraphics[page=13, trim=70 70 70 70,clip,width=1\linewidth]{Schine-suppinfo_Revised_3.pdf}
\includegraphics[page=14, trim=70 70 70 70,clip,width=1\linewidth]{Schine-suppinfo_Revised_3.pdf}
\includegraphics[page=15, trim=70 70 70 70,clip,width=1\linewidth]{Schine-suppinfo_Revised_3.pdf}
\includegraphics[page=16, trim=70 70 70 70,clip,width=1\linewidth]{Schine-suppinfo_Revised_3.pdf}
\includegraphics[page=17, trim=70 70 70 70,clip,width=1\linewidth]{Schine-suppinfo_Revised_3.pdf}
\includegraphics[page=18, trim=70 70 70 70,clip,width=1\linewidth]{Schine-suppinfo_Revised_3.pdf}
\includegraphics[page=19, trim=70 70 70 70,clip,width=1\linewidth]{Schine-suppinfo_Revised_3.pdf}
\includegraphics[page=20, trim=70 70 70 70,clip,width=1\linewidth]{Schine-suppinfo_Revised_3.pdf}
\includegraphics[page=21, trim=70 70 70 70,clip,width=1\linewidth]{Schine-suppinfo_Revised_3.pdf}
\includegraphics[page=22, trim=70 70 70 70,clip,width=1\linewidth]{Schine-suppinfo_Revised_3.pdf}
\includegraphics[page=23, trim=70 70 70 70,clip,width=1\linewidth]{Schine-suppinfo_Revised_3.pdf}
\includegraphics[page=24, trim=70 70 70 70,clip,width=1\linewidth]{Schine-suppinfo_Revised_3.pdf}
\includegraphics[page=25, trim=70 70 70 70,clip,width=1\linewidth]{Schine-suppinfo_Revised_3.pdf}
\includegraphics[page=26, trim=70 70 70 70,clip,width=1\linewidth]{Schine-suppinfo_Revised_3.pdf}
\includegraphics[page=27, trim=70 70 70 70,clip,width=1\linewidth]{Schine-suppinfo_Revised_3.pdf}
\includegraphics[page=28, trim=70 70 70 70,clip,width=1\linewidth]{Schine-suppinfo_Revised_3.pdf}

\end{document}